\documentclass[review]{elsarticle}

\usepackage{lineno,hyperref}
\modulolinenumbers[5]

\journal{Journal of \LaTeX\ Templates}

\begin{document}

\begin{frontmatter}

\title{Some remarks on the current issues in nucleon spin structure study}

\author{Israel Weimin Sun\fnref{*}}
\address{School of Physics, Nanjing University, Nanjing~210093, the People's Republic of China}
\fntext[*]{sunwm@nju.edu.cn}

\begin{abstract}
I give some personal remarks on some current issues in the nucleon spin structure study. At an elementary level I propose a new angular momentum separation for the massless
Dirac field in a free theory which mimics the usual free photon angular momentum separation pattern in Coulomb gauge. In connection with this construction I introduce
a somewhat idiosyncratic formalism in a free massless Dirac theory which I call "dressed axial $U(1)_A$ symmetry". I show that this new "fermion spin operator", which is more correctly called "helicity vector operator", can be incorporated into this new symmetry pattern in a natural way. This set of "dressed axial vector current" and its corresponding charges show an interesting internal structure and may be useful in a broader physical context. I then discuss the case of the QED model with a massless Dirac fermion. In the case of covariant quantization, I give a new and correct proof that the additional term in the total angular momentum operator stemming from the gauge-fixing term does not contribute at the level of physical matrix elements, which is discussed in the relevant literature. At the level of asymptotic fields I construct the helicity vector operator of the QED theory which actually coincides with the usual projection of the total angular momentum operator when acting on a beam of collinearly moving free particles. I then consider the QCD model with only two massless light quarks. In this context I discuss the approximate concept of "asymptotic quark and gluon fields" which is relevant to the usual parton model picture and propose to use the asymptotic "quark helicity vector" operator to describe the quark helicity contribution to an IMF proton. Finally, I give some remarks on the concept of IMF itself, which show that this very concept should be understood with some reservation from a rigorous mathematical consideration. 
\end{abstract}

\begin{keyword}
\texttt{}gauge field\sep angular momentum\sep nucleon spin structure  
\end{keyword}

\end{frontmatter}

\section{Introduction: an angular momentum separation for a free massless fermion theory}

The study of momentum and angular momentum separation problem in a gauge field system has gained much progress in the last ten years 
in the context of nucleon spin structure research due to the work of Chen ${\it et al.}$ 
\cite{Chen} in 2008. Many new developments emerge later on, which are nicely summarized in two recent review articles \cite{LeaderLorce, Wakamatsu}. 
However, there still remain some interesting issues which deserve further investigations. In this letter I shall present some personal remarks on the various
problems in these topics. First, let me begin with the simplest case, a free massless Dirac field system. 

For a free massless Dirac field, one has a standard split of the total angular momentum
\begin{eqnarray}
\nonumber {\bf J}_{free}&=& \int d^3x \big(\psi^\dagger {\bf x}\times \frac{{\bf \nabla}}{i}\psi+\psi^\dagger \frac{{\bf \Sigma}}{2}\psi\big) \\
&=& {\bf L}_{free}+{\bf S}_{free},
\end{eqnarray}
where both ${\bf L}_{free}$ and ${\bf S}_{free}$ satisfy the standard $SU(2)$ algebra and are usually regarded as the "standard" definition of orbital and spin angular momentum
operator of a Dirac field. However, for a free photon, one also has a standard separation of the total angular momentum operator in the Coulomb gauge
\begin{eqnarray}
\nonumber {\bf J}_{em}&=&\int d^3x \big(E^i_\perp {\bf x}\times{\bf \nabla} A^i_\perp+{\bf E}_\perp \times {\bf A}_\perp \big)\\
&=& {\bf L}_{\gamma free}+{\bf S}_{\gamma free},
\end{eqnarray}
where the "spin operator" ${\bf S}_{\gamma free}$ satisfies an unusual commutator ${\bf S}_{\gamma free}\times{\bf S}_{\gamma free}=0$. Physically, this phenomenon stems from the
massless nature of the photon. Then, a natural question arises: does the same thing also hold for a massless Dirac particle case?

In fact, one can write down a new angular momentum separation 
\begin{eqnarray}
\nonumber {\bf J}_{free}&=& \int d^3x \big(\psi^\dagger {\bf x}\times \frac{{\bf \nabla}}{i}\psi+\cdots+\psi^\dagger \frac{{\bf \Sigma}}{2}\cdot \frac{-i{\bf \nabla}}{\sqrt{-{\bf \nabla}^2}}\frac{-i{\bf \nabla}}{\sqrt{-{\bf \nabla}^2}}\psi\big) \\
&=& {\bf L}'_{free}+{\bf S}'_{free}.
\end{eqnarray}
One can readily check that both ${\bf L}'_{free}$ and ${\bf S}'_{free}$ are conserved (whereas ${\bf L}_{free}$ and ${\bf S}_{free}$ are not separately), and the ${\bf S}'_{free}$ also satisfies a commutator ${\bf S}'_{free}\times{\bf S}'_{free}=0$, which is quite similar to the photon spin case.

This ${\bf S}'_{free}$ has an explicit expansion in terms of the free fermion creation and annihilation operators in the helicity basis
\begin{equation}
{\bf S}'_{free}=\int \frac{d^3 p}{(2\pi)^3 2 E_{{\bf p}}}\sum_{s=\pm\frac{1}{2}}  \frac{1}{2}\frac{{\bf p}}{|{\bf p}|}\epsilon(s)(b^\dagger (p,s)b(p,s)- d^\dagger (p,s)d(p,s)).
\end{equation}
Because of this structure ${\bf S}'_{free}$ actually measures the helicity of a massless fermion times the associated unit vector ${\bf p}/|{\bf p}|$, hence I call it the "helicity vector". Due to the same reason, the free "photon spin operator" ${\bf S}_{\gamma free}$ should also be called the "helicity vector" of the free electromagnetic field. 

To unravel the internal structure of this formalism, let us also look at the helicity operator itself which reads
\begin{eqnarray}
\nonumber h &=& \int d^3x \psi^\dagger \frac{{\bf \Sigma}}{2}\cdot \frac{-i{\bf \nabla}}{\sqrt{-{\bf \nabla}^2}}\psi
= \int \frac{d^3 p}{(2\pi)^3 2 E_{{\bf p}}}\sum_{s=\pm\frac{1}{2}} \frac{1}{2} \epsilon(s)(b^\dagger (p,s)b(p,s)- d^\dagger (p,s)d(p,s)).
\end{eqnarray}
Now, let me start from the classical massless Dirac equation $i\gamma^\mu \partial_\mu \psi=0$ to obtain
\begin{equation}
{\bf \Sigma}\cdot (-i {\bf \nabla})\psi=\gamma^5 i\frac{\partial}{\partial t}\psi,
\end{equation}
which implies
\begin{equation}
{\bf \Sigma}\cdot  \frac{{\bf P}}{|{\bf P}|}\psi_{plane~wave}=\gamma^5 \epsilon(P^0) \psi_{plane~wave}.
\end{equation}
Then, by the substitution $(P^0,{\bf P})\rightarrow (i\frac{\partial}{\partial t},-i{\bf \nabla})$ for a plane wave state,
one can establish a curious identity:
\begin{equation}
\frac{{\bf \Sigma}}{2}\cdot \frac{-i{\bf \nabla}}{\sqrt{-{\bf \nabla}^2}}\psi_{clasical/operator}=\frac{1}{2}\gamma^5 \epsilon (i\frac{\partial}{\partial t})\psi_{clasical/operator},
\end{equation}
which holds for both classically on-shell $\psi_{clasical}(x)$ and quantum Dirac field operator $\psi_{operator}(x)$. With this at hand, one sees immediately
\begin{equation}
h=\int d^3x \psi^\dagger \frac{{\bf \Sigma}}{2}\cdot \frac{-i{\bf \nabla}}{\sqrt{-{\bf \nabla}^2}}\psi=\int d^3x \psi^\dagger \frac{1}{2}\gamma^5 \epsilon (i\frac{\partial}{\partial t})\psi.
\end{equation}

At this point let me introduce a general type of current which I call "dressed axial vector current"
\begin{equation}
j^{\mu 5}_f(x)={\bar \psi}\gamma^\mu \gamma^5 f(i\frac{\partial}{\partial x^\rho})\psi,
\end{equation}
where $f(\cdot)$ is an arbitrary real valued function. Using the on-shell Dirac equation, it is very easy to check that it is conserved
\begin{equation}
\partial_\mu j^{\mu 5}_f=\partial_\mu{\bar \psi}\gamma^\mu \gamma^5 f(\cdot)\psi+{\bar \psi}\gamma^\mu \gamma^5 f(\cdot)\partial_\mu\psi=0.
\end{equation}
Classically, such a dressed axial vector current is actually connected with the so-called  "dressed $U(1)_A$ rotation"
\[
\left \{
\begin{array}{c}
\psi(x)\rightarrow e^{i\theta \gamma^5 f(i\frac{\partial}{\partial x^\rho})}\psi(x) \\
{\bar \psi}(x)\rightarrow {\bar \psi}(x) e^{i\theta \gamma^5 f(\stackrel{\longleftarrow}{-i\frac{\partial}{\partial x^\rho}})}
\end{array}
\right.
\]
which can be readily shown to be a symmetry of the free massless Dirac field theory. In fact, under an infinitesimal transformation
\[
\left \{
\begin{array}{c}
\delta \psi=i\delta \theta \gamma^5 f(i\frac{\partial}{\partial x^\rho})\psi \\
\delta {\bar \psi}=i\delta \theta f(-i\frac{\partial}{\partial x^\rho}){\bar \psi} \gamma^5
\end{array}
\right.
\]
one easily verifies
\begin{eqnarray}
\nonumber \delta \mathcal{L} &\sim& i \delta \theta\big(  f(-i\frac{\partial}{\partial x^\rho}){\bar \psi}\gamma^5\gamma^\mu \partial_\mu \psi+
{\bar \psi}\gamma^\mu \gamma^5 f(i\frac{\partial}{\partial x^\rho})\partial_\mu \psi \big)\\
\nonumber &=& i \delta \theta\big( {\bar \psi}\gamma^5\gamma^\mu f(i\frac{\partial}{\partial x^\rho})\partial_\mu \psi+
{\bar \psi}\gamma^\mu \gamma^5 f(i\frac{\partial}{\partial x^\rho})\partial_\mu \psi \big) +\rm{total~divergence~term}\\
&=& \rm{total~divergence~term},
\end{eqnarray}
where the "derivative moving pattern" is apparent, e.g., for a monomial $f(\cdot)=i\frac{\partial}{\partial x^\alpha}i\frac{\partial}{\partial x^\beta}\cdots i\frac{\partial}{\partial x^\kappa}$. The corresponding conserved charge is easily constructed
\begin{equation}
Q^5(f)=\int d^3x  \psi^\dagger(x) \gamma^5 f(i\frac{\partial}{\partial x^\rho})\psi(x),
\end{equation}
which depends functionally on the real valued function $f(\cdot)$.

Now, let us come back to the helicity or helicity vector. The helicity operator $h$ is constructed using the function $f(\cdot)=\epsilon(i\frac{\partial}{\partial t})$ which is
not a smooth one. One can approximate it with a family of smooth (in fact real analytic) functions. One introduces the standard Gaussian distribution $\delta_\alpha (\tau)=\frac{1}{\sqrt{\pi}\alpha}e^{-\tau^2/\alpha}$ which tends weakly to $\delta(\tau)$ and then defines $\theta_\alpha (u)=\int^{u}_{-\infty}~d\tau \delta_\alpha (\tau)$.
With this one obtains $\epsilon_\alpha (u)=\theta_\alpha (u)-\theta_\alpha (-u)$ whose weak limit is $\epsilon(u)$. Then, one can define
\begin{equation}
h=\lim_{\alpha\rightarrow 0^{+}}\int d^3x \psi^\dagger \frac{1}{2}\gamma^5 \epsilon_\alpha (i\frac{\partial}{\partial t})\psi.
\end{equation}
The helicity vector can be constructed in the following way. One introduces a dressed current ${\bar \psi}\gamma^\mu \gamma^5\frac{1}{2}\epsilon (i\frac{\partial}{\partial t})
\frac{-i{\bf \nabla}}{\sqrt{-{\bf \nabla}^2}}\psi$ whose conserved charge is just the helicity vector
\begin{equation}
{\bf S}'=\int d^3x \psi^\dagger \frac{1}{2}\gamma^5 \epsilon (i\frac{\partial}{\partial t})\frac{-i{\bf \nabla}}{\sqrt{-{\bf \nabla}^2}}\psi=\int d^3x \psi^\dagger \frac{{\bf \Sigma}}{2}\cdot \frac{-i{\bf \nabla}}{\sqrt{-{\bf \nabla}^2}}\frac{-i{\bf \nabla}}{\sqrt{-{\bf \nabla}^2}}\psi.
\end{equation}
If one likes, he or she could also define the helicity vector as the weak limit
\begin{equation}
{\bf S}'=\lim_{\alpha\rightarrow 0^{+}}\int d^3x \psi^\dagger \frac{1}{2}\gamma^5 \epsilon_\alpha (i\frac{\partial}{\partial t})\frac{-i{\bf \nabla}}{\sqrt{-{\bf \nabla}^2}}\psi.
\end{equation}

In this context I would like to dwell on the algebraic properties of such a family of quantum conserved charges in the free field theory. First of all, a generic charge $Q^5(f)$ is
an hermitian one on the free fermion Fock space. In fact, a direct hermitian operation yields
\begin{eqnarray}
\nonumber (Q^{5}(f))^\dagger &=& \int d^3x f(-i\frac{\partial}{\partial x^\rho})\psi^\dagger(x)\gamma^5 \psi(x) \\
&=& \int d^3x \psi^\dagger(x)\gamma^5f(i\frac{\partial}{\partial x^\rho}) \psi(x),
\end{eqnarray}
which coincides exactly with the original $Q^5(f)$. Such a reasoning depends on a formal "integration by parts". This operation is legitimate for the spatial partial derivative case but it seems to be in jeopardy for a time derivative $\frac{\partial}{\partial t}$. Fortunately, the on-shell-ness of the quantum Dirac field operator rescues everything. To see this, one first notes that the so-called "doubly dressed axial-vector current"
\begin{equation}
j^{\mu 5}_{(f_1,f_2)}(x)=f_1(i\frac{\partial}{\partial x^\rho}){\bar \psi}\gamma^\mu \gamma^5 f_2(i\frac{\partial}{\partial x^\rho})\psi
\end{equation}
is also conserved. As a consequence, the corresponding charge operator $Q^5(f_1,f_2)$ is independent of time
\begin{equation}
\frac{d Q^5(f_1,f_2)}{dt}=\int d^3x \frac{\partial}{\partial t}\big( f_1(\cdot)\psi^\dagger(x) \gamma^5 f_2(\cdot)\psi(x)\big)=0,
\end{equation}
which shows that a formal "integration by parts" w.r.t. $\frac{\partial}{\partial t}$ is actually valid. Then, a direct computation using the fundamental anticommutation relations of a free Dirac field gives
\begin{equation}\label{generator}
[Q^5(f),\psi(x)]=-\gamma^5 f(i\frac{\partial}{\partial x^\rho})\psi(x),
\end{equation}
which verifies the status of $Q_f^5$ as the generator of the dressed $U(1)_A$ rotation of the Dirac field. Using (\ref{generator}) one easily finds
\begin{equation}
[Q^5(f),Q^5(g)]=0.
\end{equation}
Thus, one actually has a commuting family of conserved hermitian charges.
Finally, one notices that the correspondence $f(\cdot) \mapsto Q^5(f)$ is an algebraic homomorphism in the sense that
\begin{equation}
Q^5(c_1f_1+c_2 f_2)=c_1Q^5(f_1)+c_2Q^5(f_2).
\end{equation}

The above analysis shows vividly that some "physical observables" such as the helicity or helicity vector can be incorporated into the general framework of dressed $U(1)_A$
symmetries and the corresponding physical outcome. In some sense this could be regarded as a type of mathematical design based on some particular purposes. Nonetheless, the
clearness and flexibility of this framework proves its usefulness for handling potentially interesting physical situations.

\section{Quantum Electrodynamics}

The free fermion field case being clear, let us turn to the case of QED, e.g., the QED of just one species of massless Dirac fermion. This simple theoretical toy model is useful
since some interesting structural points already show up in such a hypothetical case.

The Lagrangian takes the form
\begin{equation}\label{QED_Lagrangian}
\mathcal{L}=-\frac{1}{4}F^2+{\bar \psi}i\gamma^\mu\partial_\mu \psi-e{\bar \psi}i\gamma^\mu\psi A_\mu,
\end{equation}
where the bare mass $m_0$ of the Dirac field vanishes. It is a standard result in the perturbative QED that to all orders of the bare electric charge a seed of the vanishing bare fermion mass $m_0=0$ will yield a vanishing pole mass $m_{pole}=0$, which implies that to all orders of perturbation theory the "physical mass" of an asymptotic free electron state
also vanishes.

In this theory the total angular momentum operator ${\bf J}$ actually depends on the quantum gauge choice. In the usual Coulomb gauge choice the angular momentum operator is of the form
\begin{equation}
{\bf J}=\int d^3x \big(\psi^\dagger {\bf x}\times \frac{{\bf \nabla}}{i}\psi+\psi^\dagger \frac{{\bf \Sigma}}{2}\psi+E^i_\perp {\bf x}\times{\bf \nabla} A^i_\perp+{\bf E}_\perp \times {\bf A}_\perp \big),
\end{equation}
where the contributions of the fermion and the photon are clearly separated.
In the standard covariant gauge quantization, the classical Lagrangian is modified as
\begin{equation}
\mathcal{L}_{cov}=\mathcal{L}-\frac{1}{2\alpha}(\partial\cdot A)^2
\end{equation}
so that manifest Lorentz covariance is preserved but the quantization procedure yields a Hilbert-Krein structure \cite{Strocchi} with an indefinite metric.
With this Lagrangian form, the angular momentum operator changes into
\begin{equation}
{\bf J}=\int d^3x \big(\psi^\dagger {\bf x}\times \frac{{\bf \nabla}}{i}\psi+\psi^\dagger \frac{{\bf \Sigma}}{2}\psi+E^i {\bf x}\times{\bf \nabla} A^i-\frac{1}{\alpha}(\partial\cdot A)({\bf x}\times{\bf \nabla})A^0+{\bf E} \times {\bf A} \big)
\end{equation}
which includes an additional contribution stemming from the gauge-fixing term. In Ref. \cite{Leader} it is pointed out that such a term does not contribute at the level of the physical matrix elements. Physically, such an intuitive conclusion is undoubtedly correct, however, the proof provided in Ref. \cite{Leader} is not without flaws. The main idea of
the proof is like this. To evaluate the physical matrix element $\langle \Phi'|\partial\cdot A~ \partial^i A^0|\Phi\rangle$ for two arbitrary physical state vectors $|\Phi'\rangle$ and $|\Phi\rangle$, one can insert "a complete set of physical states" between the two operators to obtain
$\sum_n\langle \Phi'|\partial\cdot A|\Phi_n\rangle \langle \Phi_n| \partial^i A^0|\Phi\rangle$, then because $\langle phys'|\partial\cdot A(x)|phys\rangle=0$, one arrives at
$\langle \Phi'|\partial\cdot A~ \partial^i A^0|\Phi\rangle=0 $.
However, this proof is flawed. This is because the relevant operators act on the whole indefinite metric Hilbert space, hence one should insert a complete set of intermediate states which also includes the contribution of nonphysical states (i.e., those state vectors with a negative norm), so the original proof in Ref. \cite{Leader} cannot go through without any difficulties.

Here, I shall give a new proof. The idea is rather simple: one can express all the interpolating field operators using
the free asymptotic fields and then consider everything on the in/out particle Fock space. For simplicity, I shall choose the $\alpha=1$ theory (the Feynman gauge case) at the unrenormalized level (a subsequent renormalization procedure will
"renormalize" such a gauge parameter but this is irrelevant for our essential discussions), and the quantum equation of motion of the interpolating field reads
\begin{equation}
\partial^2 A^\mu=j^\mu=e {\bar \psi}\gamma^\mu \psi.
\end{equation}
Then, a standard formal process will establish
\begin{equation}\label{interpolation}
A^\mu(x)=\sqrt{Z_3}~A^\mu_{in}(x)+\int d^4y ~D_{ret}(x-y)j^\mu(y),
\end{equation}
where everything has its standard meaning. With this at hand, let us consider the physical matrix elements $\langle \Phi'|\partial\cdot A~ \partial^i A^0|\Phi\rangle$. First, using (\ref{interpolation}) together with the current conservation condition $\partial_\mu j^\mu=0$ gives $\partial\cdot A=\sqrt{Z_3}~ \partial\cdot A_{in}$, and one also obtains
\begin{equation}\label{interpolation2}
\partial^i A^0(x)=\sqrt{Z_3}~\partial^i A^0_{in}(x)+\int d^4y ~\partial^i D_{ret}(x-y)j^0(y),
\end{equation}
then, because the electromagnetic current operator $j^\mu=e {\bar \psi}\gamma^\mu \psi$ is a gauge-invariant one, the action of the "interaction part" in (\ref{interpolation2}) on the physical state $|\Phi\rangle$ will produce a new physical state vector, which does not contribute to the matrix element $\langle \Phi'|\partial\cdot A~ \partial^i A^0|\Phi\rangle$. Consequently, what remains is a "free part" $\langle \Phi'|\partial\cdot A~ \partial^i A^0|\Phi\rangle=Z_3 \langle \Phi'|\partial\cdot A_{in}~ \partial^i A^0_{in}|\Phi\rangle$, which could then be analyzed on the free in-state Fock space. According to the usual asymptotic completeness hypothesis, the total Hilbert
space of the interacting theory is actually isomorphic to the free incoming photon/fermion Fock space, hence in a purely mathematical sense one can identify the arbitrary physical
state vectors $|\Phi\rangle$ and $|\Phi'\rangle$ appearing in the relevant matrix elements as a pair of "free physical states" in the in-state Fock space.  Then I will show that
the matrix element $\langle \Phi'|\partial\cdot A_{in}~ \partial^i A^0_{in}|\Phi\rangle$ actually vanishes.

The reasoning is as follows. On the in-state Fock space, one has the standard operator expansion
\begin{equation}\label{a}
A^0_{in}(x)=\int \frac{d^3 k}{(2\pi)^3 2\omega_k} \big( a^{(0)}(k)e^{-i k\cdot x}+ a^{(0)\dagger}(k)e^{i k\cdot x} \big),
\end{equation}
\begin{equation}\label{b}
i\partial\cdot A_{in}(x)=\int \frac{d^3 k}{(2\pi)^3 2\omega_k}|{\bf k}|\big(L(k)e^{-i k\cdot x} - L^\dagger(k)e^{i k\cdot x}\big),
\end{equation}
where the $L(k)=a^{(0)}(k)-a^{(3)}(k)$ and the physical in-state is identified by the Gupta-Bleuler condition: $\partial\cdot A_{in}^{(+)}(x)|phys\rangle=0$ (which is actually
the same as the GB condition $\partial\cdot A^{(+)}(x)|phys\rangle=0$ in the interacting theory). Then, one has
\begin{equation}
\langle \Phi'|\partial\cdot A_{in}~ \partial^i A^0_{in}|\Phi\rangle=\langle \Phi'|\partial\cdot A_{in}^{(+)}~ \partial^i A^0_{in}|\Phi\rangle=\langle \Phi'|\partial\cdot A_{in}^{(+)}~ \partial^i A^{0~(-)}_{in}|\Phi\rangle,
\end{equation}
where the last step follows from $\partial\cdot A_{in}^{(+)}~ \partial^i A^{0~(+)}_{in}|\Phi\rangle= 0$. Now, one inserts the expansion (\ref{a}) and (\ref{b}) into the relevant
matrix element and finds
\begin{eqnarray}
\nonumber && \langle \Phi'|\partial\cdot A_{in}^{(+)}(x)~ \partial^i A^{0~(-)}_{in}(x)|\Phi\rangle  \\
\nonumber &=& \int \frac{d^3 k~d^3 q}{(2\pi)^6 2\omega_k 2\omega_q}|{\bf k}|q^i \langle \Phi'|L(k)a^{(0)\dagger}(q)|\Phi\rangle e^{i(q-k)\cdot x} \\
\nonumber &=& \int \frac{d^3 k~d^3 q}{(2\pi)^6 2\omega_k 2\omega_q}|{\bf k}|q^i \langle \Phi'|[L(k),a^{(0)\dagger}(q)]|\Phi\rangle e^{i(q-k)\cdot x}  \\
\nonumber &=& -\int \frac{d^3 k}{(2\pi)^3 }\frac{k^i}{2}\langle \Phi'|\Phi\rangle \\
&=& 0,
\end{eqnarray}
which verifies $\langle \Phi'|\partial\cdot A~ \partial^i A^0|\Phi\rangle=0$. Therefore, the results discovered in Ref. \cite{Leader} are in fact correct.

Now, let us come back to the main line of our presentation. If one takes, for instance, the Coulomb gauge, one can similarly define the "helicity vector" for the fermion and the photon, respectively
\begin{equation}
{\bf S}'_f=\int d^3x \psi^\dagger \frac{{\bf \Sigma}}{2}\cdot \frac{-i{\bf \nabla}}{\sqrt{-{\bf \nabla}^2}}\frac{-i{\bf \nabla}}{\sqrt{-{\bf \nabla}^2}}\psi,
\end{equation}
\begin{equation}
{\bf S}'_\gamma=\int d^3x {\bf E}_\perp \times {\bf A}_\perp.
\end{equation}
However, in the interacting field case, these two "helicity vectors" are not conserved. But at the level of asymptotic fields one could introduce the corresponding free helicity vector operators
\begin{eqnarray}
\nonumber {\bf S}'_{in/out}&=&\int d^3x \big(\psi^\dagger_{in/out} \frac{{\bf \Sigma}}{2}\cdot \frac{-i{\bf \nabla}}{\sqrt{-{\bf \nabla}^2}}\frac{-i{\bf \nabla}}{\sqrt{-{\bf \nabla}^2}}\psi_{in/out}+{\bf E}_{\perp in/out}\times {\bf A}_{\perp in/out}\big) \\
&=& {\bf S}'_{f ~in/out}+{\bf S}'_{\gamma ~in/out},
\end{eqnarray}
which describe the "helicity vector" of free asymptotic fermions and photons and are conserved. At the formal LSZ level, one has a relation
\begin{equation}
{\bf S}'_{in} =S~{\bf S}'_{out} S^{-1},
\end{equation}
where the $S$ stands for the S-matrix of the underlying theory.

In the case of a beam of collinear incoming/outgoing free particles which I denote as $|\psi_{colinear},in/out\rangle$, it is obvious that the "helicity vector operator" ${\bf S}'_{in/out}$ can be identified as the usual projection of the total angular momentum operator of the full theory
\begin{equation}
{\bf S}'_{in/out}|\psi_{colinear},in/out\rangle= \frac{{\bf J}\cdot{\bf P}}{|{\bf P}|}\frac{{\bf P}}{|{\bf P}|}|\psi_{colinear},in/out\rangle,
\end{equation}
which shows that in this specific subspace the ${\bf S}'_{in/out}$ operator has a "gauge-invariant" meaning.

\section{Quantum Chromodynamics}

Now let us turn to the case of QCD with two massless $u$ and $d$ quarks. This is a toy model which differs from real QCD, but is appropriate for our purposes. In such a hypothetical world there should exist two bound states, the proton and the neutron, which is stable w.r.t. strong interactions. The real mass of the nucleon should be due to $\chi$SB in the underlying theory. Then, the angular momentum of the system reads (ignoring the gauge choice issues)
\begin{equation}
{\bf J}_{QCD}=\int d^3x \big(\psi^\dagger {\bf x}\times \frac{{\bf \nabla}}{i}\psi+\psi^\dagger \frac{{\bf \Sigma}}{2}\psi+E^{ai} {\bf x}\times{\bf \nabla} A^{ai}+{\bf E}^{a} \times {\bf A}^{a} \big).
\end{equation}
The real problem of nucleon spin structure is of course a bound-state problem. One should remember that, from a purely axiomatic QFT point of view, the stable nucleon state is an isolated point in the mass spectrum of QCD theory, and should be regarded as a separate composite particle with its own asymptotic in/out fields. This is rather different from the case of QED where no stable bound state exists (the only such "bound state", i.e., the positronium, is actually unstable w.r.t. electromagnetic interactions), and one is confronted
with a true bound state problem. Nevertheless, if one introduces a stable proton into the QED theory, then the Hamiltonian of the system will have a stable bound state,
the hydrogen atom, which is orthogonal to all asymptotic scattering states. In this case one would have a two-sided problem: an internal spin structure one for the proton
(please remember in that case one is studying the spin structure of an almost static proton state, not a proton observed in the IMF), and an electromagnetic angular momentum partition problem in the hydrogen atom.

Let me come back to the QCD theory and the bound state nucleon. In the usual parton model picture, an nucleon, for instance, the proton, which is observed in the IMF, could be regarded, rather approximately, as a beam of almost free and collinearly moving partons, the colored quarks and gluons.
In connection with this point I would like to give some remarks on the concept of asymptotic fields in the QCD theory. It is apparent that because of the color confinement mechanism, there are no true asymptotic quark/gluon fields which are defined in the whole $\mathbf{R}^3$ region. However, within the parton model picture, one could introduce a (frankly speaking, rather approximate and indefinite) concept of "asymptotic quark/gluon fields". This concept is based on the intuitive idea that when an actual physical process could be clearly separated into two stages: a near distance (and short time) one in which the quark/gluon particles interact sufficient weakly so that they could be effectively taken to be free ones, and a subsequent one in which confinement effects bind the colored particles into color singlet hadrons, one could "define" some sort of "asymptotic quark/gluon fields" within such a specifically chosen finite spacetime region. This concept of "asymptotic colored field" must be an approximate and spacetime-region-dependent one which is valid in a rather limited sense. On the large-scale regions, which are effectively the whole $\mathbf{R}^3$, one can define the usual asymptotic field operator of the colorless composite hadrons which have a finite size by its very formation.

Within this intuitive picture, a fast-moving proton in the parton model will be expanded as a global colorless combination of in/out quarks/gluons which are moving collinearly in the parent hadron direction. According to this approximate idea of "asymptotic quark/gluon field", one can introduce the asymptotic quark/gluon helicity vector operator acting on the IMF proton state, for instance, the quark helicity one
\begin{equation}
{\bf S}'_{q ~in/out}=\int d^3x \big(\psi^\dagger_{in/out} \frac{{\bf \Sigma}}{2}\cdot \frac{-i{\bf \nabla}}{\sqrt{-{\bf \nabla}^2}}\frac{-i{\bf \nabla}}{\sqrt{-{\bf \nabla}^2}}\psi_{in/out}\big),
\end{equation}
which measures the quark helicity contribution to the proton. 
It should be noted that in such a situation either ${\bf S}'_{q~in}$ or ${\bf S}'_{q~out}$ can be used. This is because the proton is a stable bound state, therefore according to the usual LSZ formalism there is no distinction between its in-state representation and out-state representation: $|{\rm proton}, in\rangle=|{\rm proton}, out\rangle$. As a consequence, one has
\begin{equation}
\langle{\rm proton}| {\bf S}'_{q ~in}|{\rm proton} \rangle=\langle {\rm proton}|S~ {\bf S}'_{q ~out}S^{-1}|{\rm proton}\rangle=\langle {\rm proton}| {\bf S}'_{q ~out}|{\rm proton}\rangle.
\end{equation}

Finally, some remarks on the IMF itself. From a physical point of view, the IMF should be regarded as some kind of limit of a sequence of Lorentz frames whose moving velocity tends to $c=1$. Mathematically, one can imagine to obtain an "infinite momentum proton" state, e.g., that moving in the third direction, by a boost acting on a static proton state:
\[
|{\rm proton}, p_z=\infty\rangle=\lim_{\eta\rightarrow \infty}e^{i\eta K_z}|{\rm proton}, {\rm static}\rangle
\]
I will argue that such a limit could not be reached, at least mathematically. In fact, in the usual axiomatic field theory, the quantum state vector space of the QCD theory should be a separable Hilbert space, on which one has defined a strongly continuous unitary representation of the ${\rm Poincar\acute{e}}$ group. To see why this is so, let me consider more closely the state vector: $e^{i\eta K_z}|{\rm proton}, {\rm static}\rangle$. First note that the boost operator $e^{i\eta K_z}$ is a unitary one which keeps the length of a state vector unchanged. Admittedly, a static proton state is an "improper state vector" whose length is infinite, however, one can build a normalized state vector by using, for example, a smearing process. Hence, without loss of correctness of the argument, one can assume the initial state vector $|\psi_0\rangle$  to be a normalized one. Then, could the vector $\lim_{\eta\rightarrow \infty}e^{i\eta K_z}|\psi_0\rangle$ really exist or not? Because the boost operator is indeed unitary, the whole family of the vectors $e^{i\eta K_z}|\psi_0\rangle$ all lie on the unit sphere of a separable Hilbert space which I call $\mathcal{H}_{QCD}$. However, the "limit vector" should not exist, because all state vectors in $\mathcal{H}_{QCD}$ should have a finite energy, while a $v=c=1$ boosted state has an infinite energy ! This should be compared with the situation of a pure spatial rotation operation, e.g., $U(\theta)=e^{-i \theta J_z}$ which does not change the energy. At first sight, this seems rather strange. Why there does not exist such a limit vector on this unit sphere? The reason is actually quite simple. We note that one is dealing with an infinite dimensional separable Hilbert space. Therefore, its "unit sphere" is not compact, which differs from the case of a finite-dimensional Euclidean space whose unit sphere is necessarily compact.
Because of this, a sequence of vectors $\{ e^{i\eta_n K_z}|\psi_0\rangle\}$ on that unit sphere does not necessarily have an accumulation point, which is just what actual physics teaches us. Hence, our conclusion is that the IMF and the associated parton model picture is necessarily an approximate and incomplete framework which has a limited physical significance.

\vskip 0.3cm
\noindent {\bf Acknowledgments}

The author thanks the financial support from the Natural Science Funds of Jiangsu Province of the People's Republic of China under Grant No. BK20151376.

\end{document}